\documentclass[RNAAS]{aastex62}

\begin{document}

\title{ASAS-SN Discovery of 4880 Bright RR Lyrae Variable Stars}

\correspondingauthor{T. Jayasinghe}
\email{jayasinghearachchilage.1@osu.edu}

\author[0000-0002-6244-477X]{T. Jayasinghe}
\affiliation{Department of Astronomy, The Ohio State University, 140 West 18th Avenue, Columbus, OH 43210, USA}

\author{C. S. Kochanek}
\affiliation{Department of Astronomy, The Ohio State University, 140 West 18th Avenue, Columbus, OH 43210, USA}
\affiliation{Center for Cosmology and Astroparticle Physics, The Ohio
State University, 191 W. Woodruff Avenue, Columbus, OH
43210}
\author{K. Z.  Stanek}
\affiliation{Department of Astronomy, The Ohio State University, 140 West 18th Avenue, Columbus, OH 43210, USA}
\affiliation{Center for Cosmology and Astroparticle Physics, The Ohio
State University, 191 W. Woodruff Avenue, Columbus, OH
43210}
\author{B. J. Shappee}
\affiliation{Institute for Astronomy, University of Hawai’i, 2680 Woodlawn Drive, Honolulu, HI 96822,USA}

\author{T. W. -S. Holoien}
\affiliation{Carnegie Fellow, The Observatories of the Carnegie Institution for Science, 813 Santa Barbara St., Pasadena, CA 91101, USA}

\author{T. A. Thompson}
\affiliation{Department of Astronomy, The Ohio State University, 140 West 18th Avenue, Columbus, OH 43210, USA}
\affiliation{Center for Cosmology and Astroparticle Physics, The Ohio
State University, 191 W. Woodruff Avenue, Columbus, OH
43210}
\author{J. L. Prieto}
\affiliation{N´ucleo de Astronom´ıa de la Facultad de Ingenier´ıa y Ciencias,
Universidad Diego Portales, Av. Ej´ercito 441, Santiago,
Chile}
\affiliation{Millennium Institute of Astrophysics, Santiago, Chile}
\author{Subo Dong}
\affiliation{Kavli Institute for Astronomy and Astrophysics, Peking
University, Yi He Yuan Road 5, Hai Dian District, China}

\author[0000-0002-2057-4015]{C. A. Britt}
\affiliation{College of Arts and Sciences, The Ohio State University, 140 West 18th Avenue, Columbus, OH 43210, USA}

\author{D. Will}
\affiliation{Department of Astronomy, The Ohio State University, 140 West 18th Avenue, Columbus, OH 43210, USA}


\section{}
The All-Sky Automated Survey for SuperNovae (ASAS-SN, \citealt{2014ApJ...788...48S, 2017PASP..129j4502K}) is the first optical survey to monitor the entire visible sky. ASAS-SN presently has $\sim100-500$ epochs of observation in the V-band to a depth of roughly 17 mag. The field of view of a single ASAS-SN camera is 4.5 deg$^2$, the pixel scale is 8\farcs0 and the FWHM is $\sim2$ pixels. Recently, we expanded our network of telescopes with the addition of three new ASAS-SN units (each with four telescopes) at CTIO (Chile), McDonald Observatory (Texas) and the South African Astrophysical Observatory (SAAO, South Africa). These new units use the SDSS g-band.

ASAS-SN has thus far focused on the detection of bright supernovae with minimal observational bias \citep{2017MNRAS.471.4966H}. However, many other transient and variable sources have been discovered during this process. These include roughly $\sim70,000$ new variable sources that have been flagged during the search for supernovae. As part of our effort to classify these variables (Jayasinghe et al. ~2018, in prep), we have developed a classification pipeline combining standard methods for identifying periodic variables with the open-source random forest classifier \textit{Upsilon} \citep{2016A&A...587A..18K}. Variables are matched to the VSX \citep{2006SASS...25...47W} and GCVS \citep{2017ARep...61...80S} catalogs to identify previously discovered variables. To verify the periods and classifications, members of the ASAS-SN team review each classification, with a minimum of two positive reviews needed to accept a new variable. 

\begin{figure}[htpb!]
\begin{center}
\includegraphics[scale=0.5,angle=0]{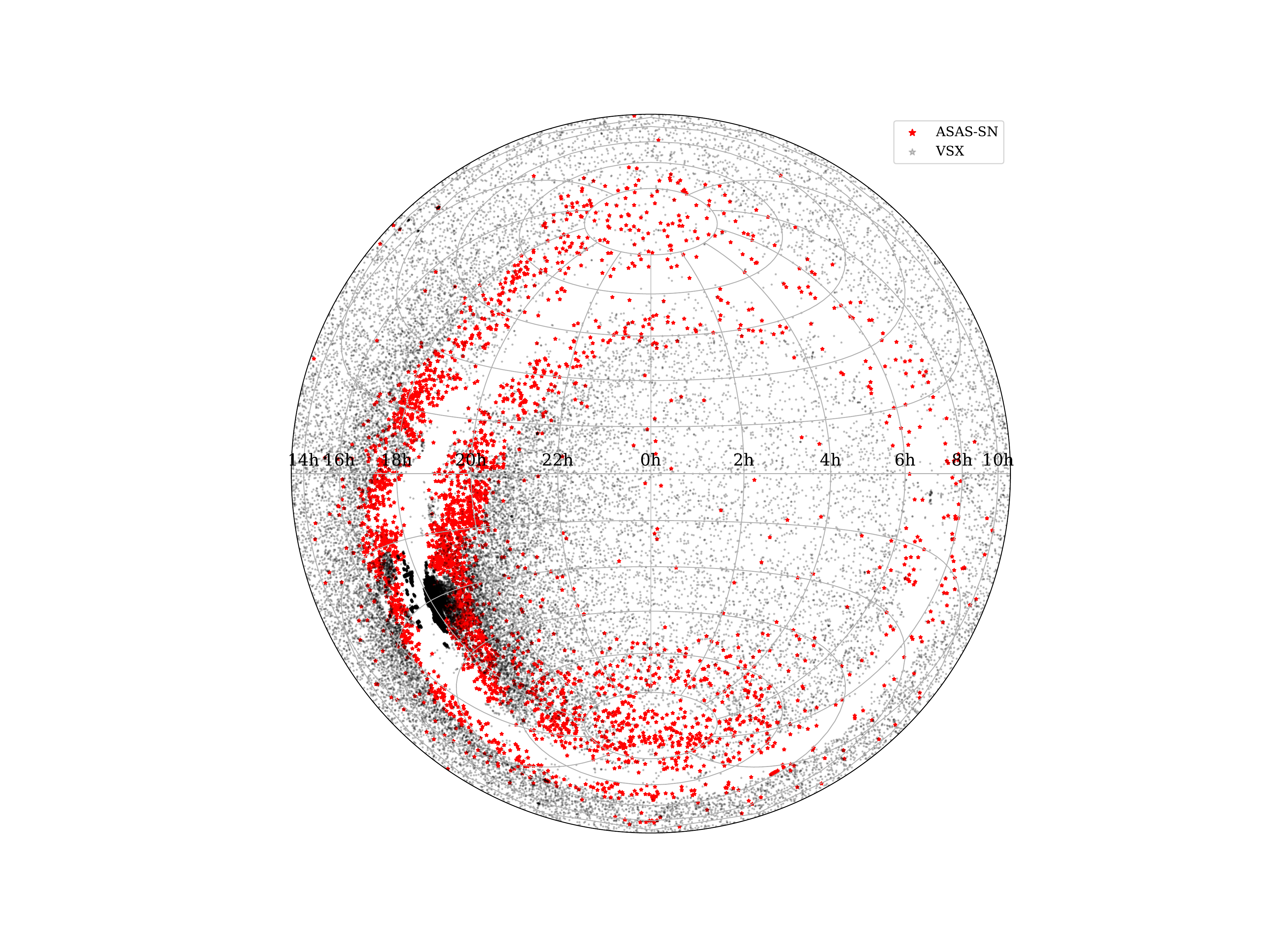}
\caption{Distribution of the 4,880 ASAS-SN (red) and 44,638 (mag $<$17) VSX (gray) RR Lyrae in Equatorial coordinates. Previous surveys avoided the Celestial poles and mid-Galactic latitudes. The high density of points near the Galactic center results from I-band detections of RR Lyrae by OGLE \citep{2014AcA....64..177S} around Baade's and Stanek's \citep{1998astro.ph..2307S} windows.\label{fig:1}}
\end{center}
\end{figure}

Here we report 4880 new RR Lyrae variables detected in the V-band. ASAS-SN's typical astrometric error is $\sim$1" and we find that 99\% of the new RR Lyrae are within 4\farcs4 of a 2MASS \citep{2006AJ....131.1163S}, UCAC4 \citep{2013AJ....145...44Z}, NOMAD \citep{2004AAS...205.4815Z},  USNOB1.0 \citep{2003AJ....125..984M} or PPMXL \citep{2010AJ....139.2440R} star of the correct magnitude. We updated the ASAS-SN coordinates to that of the cataloged star. Our RR Lyrae sample consists of 4433 RRab, 446 RRc and 1 RRd variables with mean magnitudes between 12.1-17.1 mag and periods from 0.22 to 0.93 days. We detect RR Lyrae over a range of amplitudes, from 0.10 to 1.83 mag, where the amplitude is defined as the difference between the 95$^{th}$ and 5$^{th}$ percentiles of the magnitudes. Figure \ref{fig:1} shows the distribution of the V$<$17 mag ASAS-SN RR Lyrae over the celestial sphere, as compared to the 44,638 RR Lyrae in the VSX catalog with magnitudes $<$17 mag. It is evident that prior surveys avoided mid Galactic latitudes and the equatorial poles.

We have submitted these RR Lyrae to the AAVSO and they are now included in the latest VSX catalog \citep{2006SASS...25...47W}. These RR Lyrae were submitted to the AAVSO prior to the public release of the Pan-STARRS1 3$\pi$ RR Lyrae catalog \citep{2017AJ....153..204S}. This data release comes as the first installment of our effort to classify variables in the ASAS-SN pipeline (Jayasinghe et al. 2018, in prep). The ASAS-SN light curves for these new variables and increasing numbers of known variables can be retrieved at the new ASAS-SN light curve server (\href{https://asas-sn.osu.edu/variables}{https://asas-sn.osu.edu/variables}).

\bibliographystyle{plainnat}

\end{document}